\documentclass[iop,twocolappendix]{emulateapj}
\usepackage{color}
\usepackage{natbib}
\usepackage{graphicx}
\usepackage{amsmath}
\usepackage{collectbox}
\usepackage[breaklinks,colorlinks,allcolors=blue]{hyperref}
\newcommand{\Msun}{\mathrm{M}_{\odot}}
\newcommand{\Lsun}{\mathrm{L}_{\odot}}

\newcommand{\ks}{K_{\mathrm{s}}}
\newcommand{\wjk}{W_{\mathrm{JK}}}
\newcommand{\logPWJK}{\log(P_{W_{\mathrm{JK}}=12})}
\newcommand{\logPWJKone}{\log(P_{1,W_{\mathrm{JK}}=12})}

\slugcomment{Accepted for publication in ApJ}

\shorttitle{Period-Luminosity Sequences of LPVs}
\shortauthors{Trabucchi et al.}

\begin{document}

\title{A new interpretation of the period-luminosity sequences \\ of long-period variables}
\email{michele.trabucchi@phd.unipd.it}

\author{Michele Trabucchi}
\affil{Dipartimento di Fisica e Astronomia Galileo Galilei
    Universit{\`a} di Padova, Vicolo dell'Osservatorio 3, I-35122 Padova, Italy}

\author{Peter R. Wood}
\affiliation{Research School of Astronomy and Astrophysics,
    Australian National University, Canberra, ACT2611, Australia}

\author{Josefina Montalb{\'a}n}
\affiliation{Dipartimento di Fisica e Astronomia Galileo Galilei
    Universit{\`a} di Padova, Vicolo dell'Osservatorio 3, I-35122 Padova, Italy}

\author{Paola Marigo}
\affiliation{Dipartimento di Fisica e Astronomia Galileo Galilei
    Universit{\`a} di Padova, Vicolo dell'Osservatorio 3, I-35122 Padova, Italy}

\author{Giada Pastorelli}
\affiliation{Dipartimento di Fisica e Astronomia Galileo Galilei
    Universit{\`a} di Padova, Vicolo dell'Osservatorio 3, I-35122 Padova, Italy}

\author{L{\'e}o Girardi}
\affiliation{Astronomical Observatory of Padova - INAF,
    Vicolo dell'Osservatorio 3, I-35122 Padova, Italy}


\begin{abstract}

Period-luminosity (PL) sequences of long period variables (LPVs)
are commonly interpreted as different pulsation modes,
but there is disagreement on the modal assignment. 
Here, we re-examine the observed PL sequences in the Large Magellanic Cloud,
including the sequence of long secondary periods (LSPs),
and their associated pulsation modes.
Firstly, we theoretically model the sequences using linear,
radial, non-adiabatic pulsation models
and a population synthesis model of the LMC red giants.
Then, we use a semi-empirical approach
to assign modes to the pulsation sequences
by exploiting observed multi-mode pulsators.
As a result of the combined approaches,
we consistently find that sequences B and C$^{\prime}$
both correspond to first overtone pulsation,
although there are some fundamental mode
pulsators at low luminosities on both sequences.
The masses of these fundamental mode pulsators are larger
at a given luminosity than the mass of the first overtone pulsators.
These two sequences B and C$^{\prime}$ are separated by a small
period interval in which large amplitude pulsation
in a long secondary period (sequence D variability) occurs, meaning that
the first overtone pulsation is not seen as the primary mode of pulsation.
Observationally, this leads to the splitting of the first overtone
pulsation sequence into the two observed sequences B and C$^{\prime}$.  
Our two independent examinations also show that
sequences A$^{\prime}$, A and C correspond to third overtone, second
overtone and fundamental mode pulsation, respectively.

\end{abstract}

\keywords{stars: AGB and post-AGB - stars: oscillations - stars: variables: general - Magellanic Clouds}


\section{Introduction}
\label{seq:intro}

Luminous red giants are known to be variable,
and their periods of observed variability lie on
well defined sequences in period-luminosity (PL) diagrams.
\citet{Wood1999} and \citet{Wood2000} identified five sequences
(labelled A, B, C, D and E) in the PL diagram of the Large Magellanic Cloud (LMC)
using observations from the MACHO project.

\citet{Ita2004} found that sequence B is in fact two sequences,
B and C$^{\prime}$, while \citet{Soszynski2004OSARG}
showed that an additional sequence exists on the short-period side of sequence A.
They labelled that sequence a$_4$, although we use here
the alternative designation A$^{\prime}$,
as in \citet{Wood2015} and references therein.
Moreover, sequences A and B, and possibly A$^{\prime}$,
each consist of three or more closely spaced sequences
\citep{Soszynski2004OSARG, Soszynski2007}.

Here, we are interested in sequences A$^{\prime}$, A, B, C$^{\prime}$ and C,
which are attributed to pulsating stars.
All these long period variables (LPVs) can be either
on their red giant branch (RGB) or asymptotic giant branch (AGB) phases,
with just a slight offset between the respective period distributions
\citep{Ita2002, KissBedding2003}.
Sequence D consists of long secondary periods (LSPs),
while sequence E is due to binary stars.
We do not discuss sequence E further in this paper.

Existing interpretations of the observed sequences
A$^{\prime}$, A, B, C$^{\prime}$ and C all assume that
these sequences correspond to distinct and adjacent radial orders of pulsation.
There are two incompatible interpretations in the literature.
\cite{Wood2015} assumed that sequence C, containing the Mira variables,
corresponded to the radial fundamental mode
so that sequences C$^{\prime}$, B, A and A$^{\prime}$ corresponded
to the radial 1st, 2nd, 3rd and 4th overtones, respectively.
On the other hand, \cite{Mosser2013} and \cite{Soszynski2007}
matched sequences B and A to the 1st and 2nd radial overtones,
meaning that sequence C$^{\prime}$ is the fundamental mode
and that sequence C has no explanation.
Note that there is a one sequence offset between the above two sets of mode assignments.

A solution to the above mode assignment incompatibility is to relax the requirement
that the adjacent observed sequences correspond to adjacent radial modes of pulsation,
where we identify radial order $n=0$ with the fundamental mode,
radial order $n=1$ with the 1st overtone (1O) mode, and so on.

Here, we explore the possibility that sequence B corresponds
to 1O mode pulsation as suggested by \cite{Mosser2013},
sequence C corresponds to radial fundamental mode pulsation as required by \cite{Wood2015},
but that sequence C$^{\prime}$ is associated with
1O or fundamental radial mode, or both these modes.
To do this, we firstly construct theoretical pulsation sequences and compare them
with observations, following which we take an independent semi-empirical approach
and examine observations of stars exhibiting multiple periods.


\section{Period-luminosity relations in a population synthesis model}
\label{sec:simulation}


\subsection{Methods}
\label{sec:simulation_methods}

We used the code \texttt{TRILEGAL} \citep{trilegal}
to compute a synthetic stellar population representative of the LMC,
assuming a constant star formation rate from ages
$\sim5\mathrm{Myr}$ to $\sim10\mathrm{Gyr}$
and the age-metallicity relation by \cite{Pagel1999}.
Such a simplifying assumption is made to eliminate discontinuities
that could be caused by periods of reduced or null star formation.
The technical details of such simulations
are discussed elsewhere (Pastorelli et al., in prep.).
Here, it suffices to recall that the simulations include single stars
in all relevant evolutionary phases,
as extracted from the \texttt{PARSEC} \citep{parsec}
and \texttt{COLIBRI} \citep{colibri} grids of stellar evolutionary tracks.
As to the AGB phase, stellar models account for the formation of carbon stars
due to the third dredge-up, as well as for
the occurrence of hot-bottom burning \citep{isoc2017}.
The simulation output includes many intrinsic stellar quantities
such as luminosity, radius, core mass, surface abundances,
and evolutionary stage, as well as the apparent photometry.

We then used the radial, non-adiabatic pulsation models
described in \citet{WoodOlivier2014} and Trabucchi et al., in prep.,
to assign periods and growth rates for
five radial modes to each simulated RGB or AGB star.
Models include detailed atomic and molecular opacities
\citep{aesopus} for a realistic description of carbon stars.
The growth rates provide an indication of the stability of pulsation modes:
a mode having a negative growth rate should be stable,
while a positive growth rate means a mode is excited.
In a given stellar model, we call the \textit{dominant}
mode the one with the largest growth rate,
and it is also expected to have the largest amplitude.


\subsection{Comparison with observations}
\label{sec:simulation_comparison}

\begin{figure*}[ht!]
\plotone{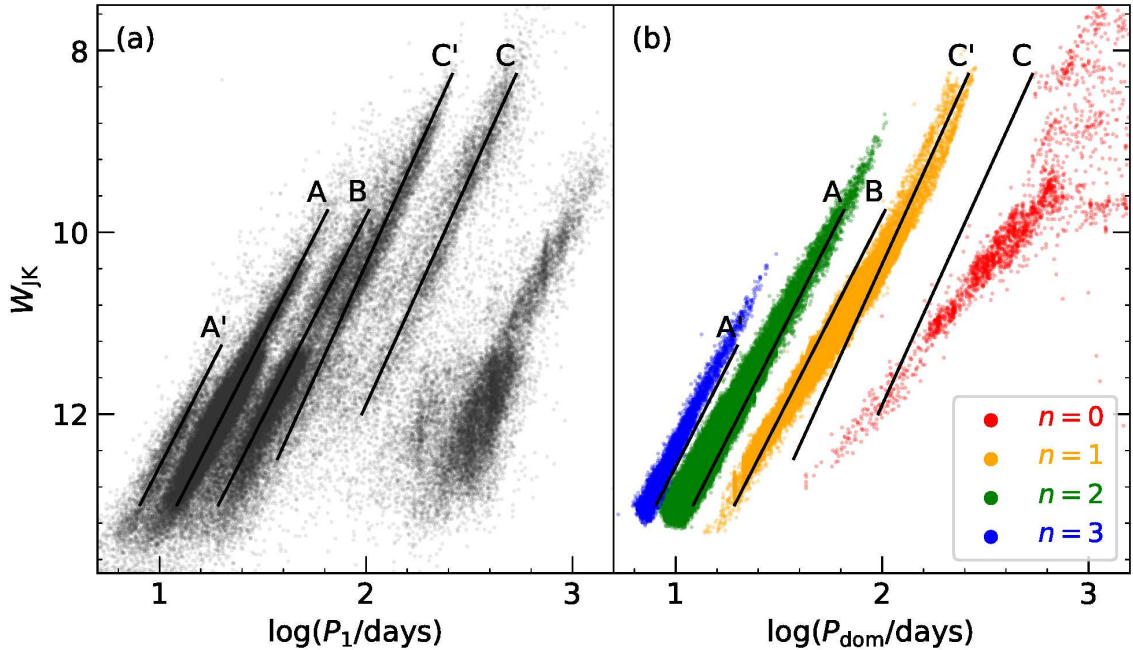}
\caption{
    Observed primary periods (panel (a)) are compared to theoretically 
    dominant periods obtained from radial, linear, non-adiabatic models (panel (b)).
    Different colours represent periods of different modes:
    red for fundamental mode, orange for 1O, green for 2O mode, blue for 3O mode.
    Solid lines provide a visual aid to identify
    the approximate location and slope of observed sequences.
\label{fig:comparison_radial_dominant}}
\end{figure*}

We compare the results with observations using data
from the OGLE III Catalogue of LPVs in the LMC \citep{Soszynski2009LMC}
combined with near-infrared photometry from the
Two Micron All Sky Survey (2MASS) Point Source Catalogue \citep{2mass}.
The comparison is made in the $\log$($P$)-$\wjk$ plane, where the Weisenheit index defined as
\begin{equation}
    \wjk = \ks - 0.686 (J - \ks)
\end{equation}
is a reddening-free measure of the luminosity.

We use periods derived from radial, non-adiabatic models,
and compare the dominant mode (the mode with the highest growth rate)
in the models to the observed primary modes in the OGLE data.
The latter are the ones associated with the highest peak
in the power spectrum, and have generally the largest amplitude.
Theoretical modes having a positive but small growth rate are not expected to be observed,
therefore only dominant modes with a growth rate larger than $0.04$ are shown,
where the growth rate is the fractional growth in amplitude per period
as defined by \citet{WoodOlivier2014}.
The comparison is shown in Fig.~\ref{fig:comparison_radial_dominant}.

We find the models to be in reasonably good agreement with observations.
The observed sequences can be explained with exactly
four radial oscillation modes, from the fundamental to the 3O mode.
The radial 4O mode, also computed for the simulation,
is never predicted to be dominant,
and is not required to reproduce the observations.

Sequences A$^{\prime}$ and A are fairly well reproduced
by radial 3O and 2O mode periods, respectively,
although in both cases the theoretical sequences are slightly offset
towards short periods with respect to observed ones.

When the radial 1O mode is dominant, it has periods extending 
across both sequences B and C$^{\prime}$.
However, there is no theoretical counterpart in the models
for the gap between the two sequences
(this will be discussed in Section~\ref{sec:gap}).

Dominant fundamental periods are distributed mainly to the right of sequence C,
and are not as a good match to the observations as are the overtones.
However, according to the models, the fundamental mode is the only mode
possibly responsible for sequence C, as no other mode can reach such long periods.
The reasons for the poor agreement between the positions
of sequence C and the positions of predicted fundamental mode pulsators 
are explored further in Section~\ref{sec:amplitudes}.


\subsection{Amplitudes and growth rates}
\label{sec:amplitudes}

\begin{figure*}[ht!]
\plotone{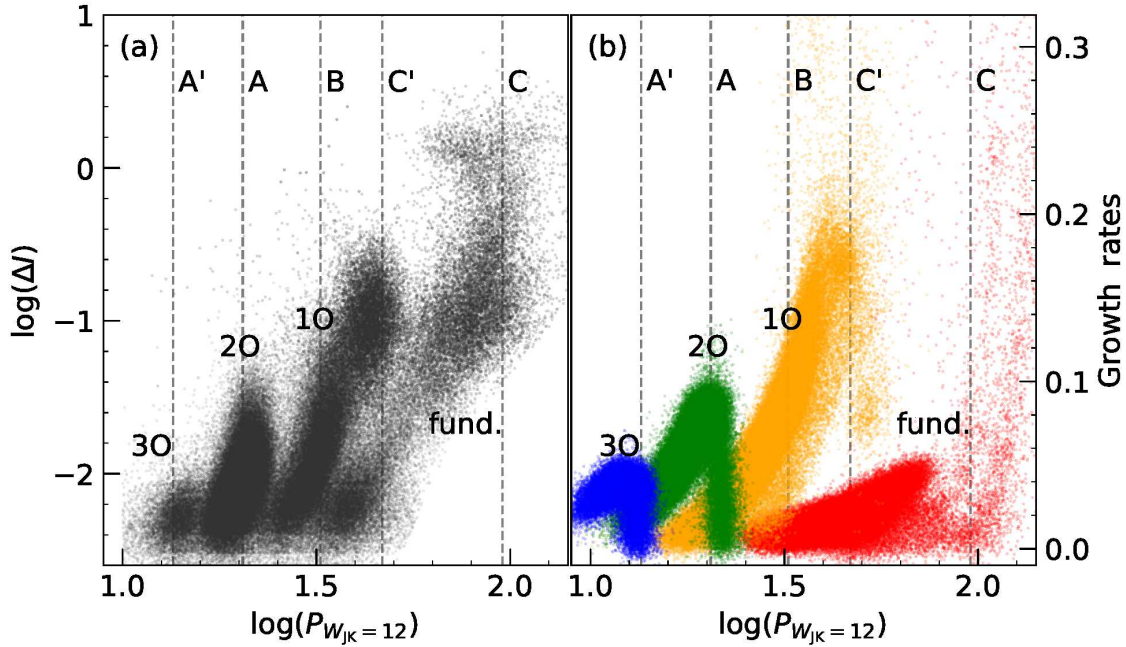}
\caption{
    \textbf{Panel (a):} logarithm of observed $I$ band amplitudes,
    for all three periods of stars in the OGLE3 catalogue, as a function of $\logPWJK$.
    The distinct PAL relations are labelled with the corresponding mode
    according to the interpretation provided here (see text).
    Periods on sequence D and longer than sequence C
    have been ignored, as they are not related to the
    normal modes of pulsation (see Section~\ref{sec:observed_multimodes}).
    \textbf{Panel (b):} theoretical growth rates for all excited modes
    up to the 3O, labelled and colour coded according to their
    radial order as in Fig.~\ref{fig:comparison_radial_dominant}.
    Vertical dashed lines in the background of both panels
    approximately identify the centre of the PL sequences.
\label{fig:amplitude_relations}}
\end{figure*}

Here, we examine observed amplitudes
to interpret LPVs in terms of different pulsation modes.
In panel (a) of Fig.~\ref{fig:amplitude_relations},
the distribution of observed amplitudes\footnote{
Amplitude data in the OGLE3 catalogue
comes rounded to the third decimal digit.
As a consequence, visual inspection of
low-amplitude distributions in log-scale is virtually impossible.
For this reason, we added random errors
with a standard deviation of $7.5\cdot10^{-4}\,\mathrm{mag}$
to amplitude data, making their distribution
visually smooth in Fig.~\ref{fig:amplitude_relations}.}
is shown against the variable $\logPWJK$ defined by
\begin{equation}\label{eq:logPWJK12}
    \logPWJK = \log(P) + \frac{\wjk - 12}{4.444} \,,
\end{equation}
which represents $\log$($P$) projected along a line parallel to 
the observed sequences until it meets the level $\wjk$~$=$~$12$.
This way, periods on a given sequence are bounded
to the same horizontal range \citep[see also][]{Wood2015}.

Similarly to what happens in the PL diagram,
where observations lie on well defined sequences,
in Fig.~\ref{fig:amplitude_relations} observed modes
are distributed into four distinct groups,
that are actually period-amplitude-luminosity (PAL) relations
(luminosity plays a role through the dependence
of $\logPWJK$ upon $\wjk$ in Eq.~\ref{eq:logPWJK12}).
At short periods ($\logPWJK\lesssim1.2$),
the cluster of observed modes of small amplitude 
is the counterpart of sequence A$^{\prime}$,
while the one with $1.2\lesssim\logPWJK\lesssim1.4$,
and extending up to larger amplitudes, corresponds to sequence A.
Towards long periods, clusters become elongated and slanted.
The one starting at $\logPWJK\simeq1.4$ with small amplitude,
corresponding to sequence B, culminates at larger amplitude
at $\logPWJK\simeq1.6$, where it is the counterpart of
the bright end of sequence C$^{\prime}$.
Finally, the faint end of sequence C$^{\prime}$
corresponds to the clump at small amplitude at
$\logPWJK\simeq1.6$, which extends to larger amplitudes
at $\logPWJK\simeq1.9$, corresponding to sequence C.
It then shows a tendency to bend back to the short
period side of sequence C at the largest amplitudes,
corresponding to Miras, whose amplitude
is limited by nonlinear effects. 

In the previous subsection,
we showed that the region occupied by the 
sequences in the PL diagram is also reproduced
by exactly four radial modes.
This suggest that the four PAL relations
can be identified with radial modes from the fundamental to the 3O.
To support this hypothesis, we use theoretical growth rates
as a proxy for mode amplitude in order to make a comparison
with the observed distribution of amplitudes,
as shown in panel (b) of Fig.~\ref{fig:amplitude_relations}.

The distribution of growth rates there exhibits
a characteristic behaviour typical of the evolution
of pulsational stability in luminous red giants
(this will be discussed in section~\ref{sec:stability_evolution},
see also Fig.~\ref{fig:gr-lgl_4modes_alphanu0.0}).  
At short periods, the 3O mode is dominant (has the largest growth rate)
and its period falls in the vicinity of sequence A$^{\prime}$.
With further evolution, its growth rate drops,
forming the vertical strip at $\logPWJK\simeq1.1$.
The 2O mode shows the same behaviour with changing $\logPWJK$.
For the 1O mode, the growth rate rises with $\logPWJK$
as for the 2O and 3O modes but the vertical strip
corresponding to the growth rates dropping to negative values is absent.
The fundamental mode shows a different pattern,
being stable or weakly unstable most of the time,
and becoming suddenly very unstable
(very large growth rates, like those at $\logPWJK\gtrsim2$).

In Fig.~\ref{fig:amplitude_relations},
the observed amplitudes seen in panel (a)
and the predicted amplitudes seen in panel (b)
describe an entirely similar pattern
as long as only overtone modes are considered.
We interpret this agreement as a validation of our hypothesis
to assign a different radial order of oscillation to each PAL relation.
It also supports the idea that sequences B and C$^{\prime}$
are both due to 1O mode pulsation.
The remaining PAL relation has periods longer than the 1O mode
and is related to sequence C, which is almost certainly due
to fundamental mode pulsation \citep{Wood2015}.
Thus, we identify it with the fundamental mode
in spite of the poor agreement of models with observations.

Such a poor agreement is a consequence of the fact
that the fundamental mode is generally well reproduced
in terms of periods, but not of driving.
We further explore this aspect in Section~\ref{sec:stability_evolution}.

\begin{figure*}[ht!]
\plotone{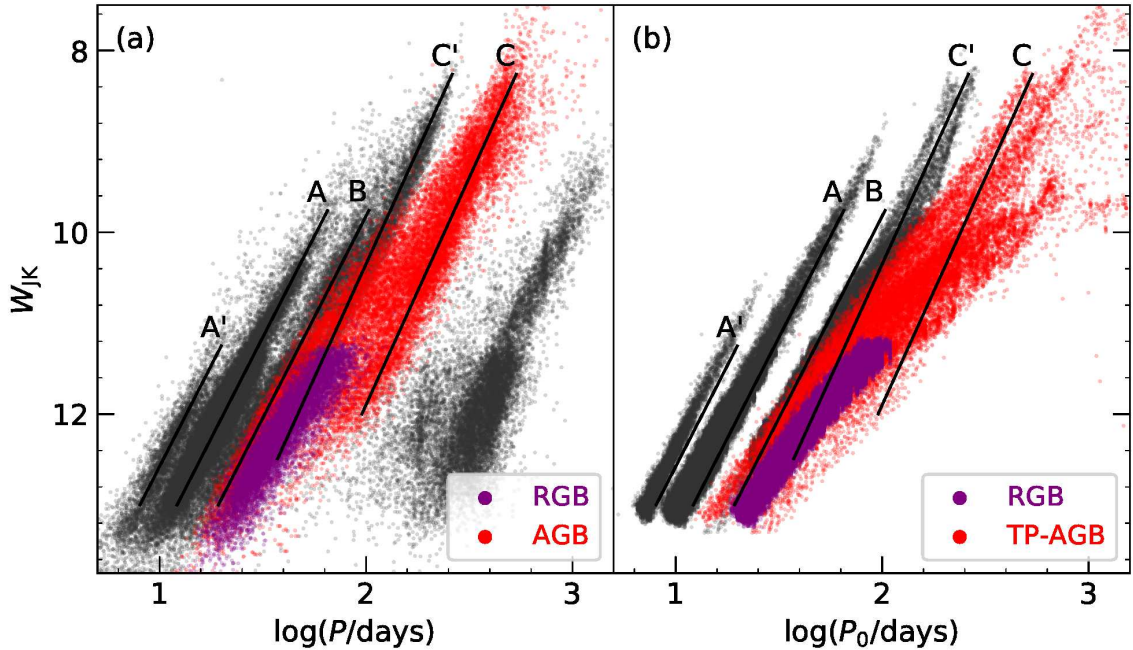}
\caption{
    \textbf{Panel (a).}
    Each star in the OGLE3 Catalogue has three observed periods:
    primary, secondary and tertiary.
    Here, all of them are shown,
    provided that they belonging to the fourth amplitude group
    in Fig.~\ref{fig:amplitude_relations} (labelled ``fund.'').
    They are shown as red or purple points if classified, respectively,
    as AGB or RGB stars (according to the OGLE evolutionary
    classification as in \cite{Soszynski2004OSARG}).
    As a reference, all primary periods are shown
    in the background as dark points (same as Fig.~\ref{fig:comparison_radial_dominant}).
    \textbf{Panel (b).} Theoretical fundamental mode periods,
    regardless of their growth rates,
    shown in red if belonging to TP-AGB models
    and in purple if belonging to RGB models.
    Dominant overtone modes are shown as dark points
    in the background for reference.
\label{fig:fundamental_mode}}
\end{figure*}

To show that the issue is limited to the modelling of driving,
and to validate our identification of PAL relations with pulsation modes,
in Fig.~\ref{fig:fundamental_mode} we examine the distribution 
of \textit{all} theoretical fundamental periods,
regardless of their growth rates.
We compare this distribution with the PAL
relation of longest period (the one labelled ``fund.''),
including primary, secondary and tertiary periods
rather than only the primary periods.
Fundamental mode periods,
both in the observations and in the models,
cover not only the region of sequence C,
but also a wide area including the the faint part
of sequences B and C$^{\prime}$.
The agreement found is a confirmation that
fundamental mode periods are generally correct,
although slightly overestimated.


\section{Semi-empirical analysis of the pulsation sequences of LPVs}
\label{sec:empirical}


\subsection{The pulsation modes associated with the observed PL sequences}
\label{sec:observed_multimodes}

\begin{figure*}[ht!]
\epsscale{1.04}
\plotone{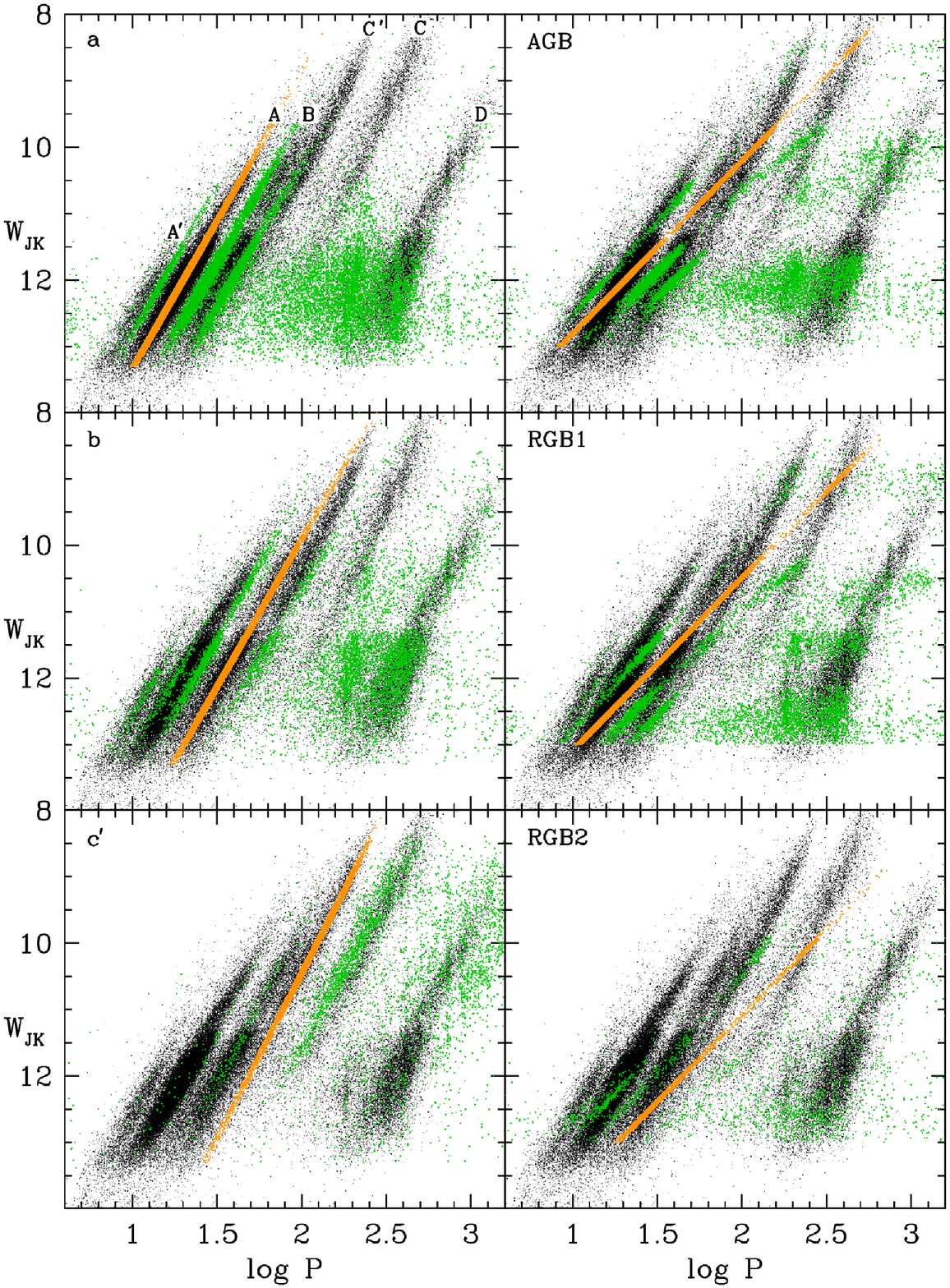}
\caption{
    Plots of $\wjk$ against $\log(P)$.
    Black points in each panel
    are the primary periods of each star.
    These points fall mostly on the six sequences labelled in panel (a).
    In the left three panels, a channel of small width centred on a sequence
    is defined and all primary periods in these channels are coloured orange.
    The secondary and tertiary periods of these stars are plotted as green points.
    In the right three panels, a sloping channel approximating an evolutionary track is defined 
    and the same colour scheme is used for primary, secondary and tertiary periods.
\label{fig:wjk-logp_ogleIII_fixed_channels}}
\end{figure*}

In this section, we examine the periods in the many multimode LPVs in the LMC.
In Fig.~\ref{fig:wjk-logp_ogleIII_fixed_channels}
we show six $\log$($P$)-$\wjk$ diagrams for LPVs in the LMC.
Black points in the background are observed primary periods,
and form the approximately parallel PL sequences
A$^{\prime}$, A, B, C$^{\prime}$ and C.  
The sequence identifications are shown on panel (a)
of Fig.~\ref{fig:wjk-logp_ogleIII_fixed_channels}.

Panel (a) of Fig.~\ref{fig:wjk-logp_ogleIII_fixed_channels} 
shows the secondary and tertiary periods (green points)
associated with stars whose primary periods (orange points)
lie is a channel down the centre of sequence A.
There are clearly three additional modes
associated with those stars\footnote{
    We ignore the broad area of green points
    on the long period side of sequence C
    which are a mixture of periods related
    to the presence of LSPs on sequence D
    and many other periods caused
    by the annual observing cycle for the OGLE data
    and possibly by convection-related variability.
}.
The most important point to note about panel (a)
is that there are only two modes (green strips)
with periods longer than the mode associated with sequence A
(sampled by the orange strip).
Assuming the longest period mode corresponds
to fundamental mode pulsation, then sequence A
must correspond to second overtone (2O) pulsation.
There is one green strip in panel (a)
with a period shorter than that of sequence A
and this must correspond to the 3O mode of pulsation.  
Sequence A$^{\prime}$, which is on the short period side of sequence A,
should therefore also correspond to the 3O mode of pulsation.
Note that, for simplicity, we consider here only radial modes
(angular degree $\ell=0$), although nonradial modes are also present in LPVs.
Indeed, dipole ($\ell=1$) and quadrupole ($\ell=2$) modes are
present within sequences A and B and possibly A$^{\prime}$,
as discussed by  \citet{Soszynski2004OSARG}, \citet{Soszynski2007} and \citet{Wood2015}.
\citet{Stello2014} analysed a sample of late K and early M type giants
on the RGB which were observed by \textit{Kepler} and they 
found their power spectra to be dominated by dipole
modes for the 2nd and higher overtones, even for the most luminous stars in their sample.
This suggests that nonradial modes may also
be dominant for the 2nd and higher overtones in LPVs in the OGLE sample,
which overlaps in luminosity the sample used by \citet{Stello2014}.
Since the dipole and quadrupole periods are longer than
the radial period of the same radial order, the dominance
of the dipole mode could help explain why the predicted 3O and 2O
periods for sequences A$^{\prime}$ and A, respectively, 
in panel (b) of Fig.~\ref{fig:comparison_radial_dominant} are slightly shorter 
than the mean observed periods for these two sequences.
We are currently making nonradial pulsation models
for LMC AGB stars in order to better understand the
observed nonradial modes and their contribution to the
period-luminosity sequences (Montalban et al., in prep).

Moving to panel (b) of Fig.~\ref{fig:wjk-logp_ogleIII_fixed_channels},
we see that for the stars whose primary period lies in sequence B,
there is only one associated mode
whose period is longer than the sequence B period.
Once again, if this longer period mode is the fundamental mode,
then the stars on sequence B which exhibit the longer period
correspond to 1O pulsaton. This is consistent
with a picture whereby the adjacent sequences A and B
are associated with adjacent radial modes.
In panel (b), there appear to be three additional modes (green strips)
shorter than the sequence B mode at low luminosities.
These would be the 4O, 3O and 2O modes
if all the stars on sequence B where 1O mode pulsators.
However, this is not the case.
In the faint part of sequence B there are in fact two groups of stars,
each with a different mass at a given luminosity.
The first group consists of the 1O mode pulsators mentioned above.
The second group consists of more massive stars
whose primary period normally belongs to sequence A$^{\prime}$
and which, by chance, have their longest period mode
coinciding with sequence B, as shown by \cite{Wood2015}.
For these stars, the mode found on sequence B is the fundamental mode
so that, at low luminosities, sequence B contains both 1O
and fundamental mode pulsators.
Normally, the latter group would have a primary period
located on sequence A${^\prime}$ (the 3O mode),
but sometimes the fundamental mode
is detected as the primary pulsation mode,
as both have rather small amplitudes.

Moving to panel (c$^{\prime}$) of Fig.~\ref{fig:wjk-logp_ogleIII_fixed_channels}, 
we see that for the stars whose primary period lies in sequence C$^{\prime}$,
there is also one associated mode whose period
is longer than the sequence C$^{\prime}$ period,
at least for AGB stars above the tip of the RGB
at $\wjk \approx 11.3$ and for some
stars (possibly AGB stars) just below the RGB tip.
This indicates that for these stars,
sequence C$^{\prime}$ corresponds to 1O pulsation,
just like the luminous stars on sequence B.
The longer period mode coincides with sequence C
or its shorter period edge,
consistent with the assumption that
sequence C is made up of fundamental mode pulsators
as argued by \citet{Wood2015}.
In addition to the sequence C$^{\prime}$ stars
exhibiting a longer secondary or tertiary period,
there are some C$^{\prime}$ stars at lower luminosities
that exhibit a shorter period coinciding with sequence B.
These are the analogs of the sequence B stars
which have their fundamental mode on
or to the long period side of sequence C$^{\prime}$.
Thus, especially at RGB luminosities,
sequence C$^{\prime}$ contains some fundamental mode pulsators.
As with sequence B, the fundamental mode pulsators on sequence C$^{\prime}$
are more massive at a given luminosity
than the 1O mode pulsators on this sequence.
Note that sequence C$^{\prime}$ is sparsely populated at low luminosities
and it is poorly defined there.

We now have a picture in which sequences B and C$^{\prime}$
are made up predominantly of 1O pulsators but
at low luminosities they also contain fundamental mode pulsators
of higher mass at a given luminosity.
Note that this is consistent with the results of Section~\ref{sec:amplitudes}
based on the identification of PAL relations
with the four lowest order pulsation modes.
In this situation, the idea that adjacent sequences
correspond to adjacent radial orders of pulsation breaks down.
Another anomaly is that sequences B and C$^{\prime}$
seem to be unusually close together in $\log(P)$.
The longer period mode associated with sequence B stars
has a period that is longer than the period of the adjacent sequence, C$^{\prime}$.
For all other sequences, the secondary and tertiary periods
form strips that are closer to their primary mode sequence
than are the adjacent primary mode sequences.
\citet{Wood2015} argued that this is because the mass of stars
at a given luminosity decreases in moving
from sequence A$^{\prime}$ through to C.
In this situation, and under the assumption that
sequences B and C$^{\prime}$ corresponded to adjacent radial orders,
\citet{Wood2015} tried to explain
the unusually close positions of sequences B and C$^{\prime}$
by assuming the sequences C$^{\prime}$ and C were radial $\ell=0$ pulsators
but that sequences A$^{\prime}$, A and B were dominated by nonradial $\ell=1$ pulsators
whose periods were slightly larger than those of the $\ell=0$ pulsators.
If we relax the condition that sequences B and C$^{\prime}$
coincide with adjacent radial orders, such a strong assumption
is not necessary anymore, and we can safely assume the presence
of dominant modes on sequences A$^{\prime}$, A and B that are due to radial pulsation
(although, as noted above, nonradial modes have a presence within these sequences).

In order to throw further light on the modes associated with the PL sequences,
in the right hand panels of Fig.~\ref{fig:wjk-logp_ogleIII_fixed_channels}
we consider sloping channels in the $\log$($P$)-$\wjk$ diagram
so that we can follow mode evolution
in passing from one sequence to the next.
Remember that the orange points are primary modes (the largest amplitude)
and the green points are the secondary and tertiary modes (smaller amplitude).

As the orange channel crosses from sequence B to sequence C$^{\prime}$
in the right hand panels of Fig.~\ref{fig:wjk-logp_ogleIII_fixed_channels},
we see that the fundamental mode, represented by the longest period green strip,
increases in period from that of sequence C$^{\prime}$ to that of sequence C.
This green strip terminates at the right hand edge of sequence C
due to the fact that very large amplitude Mira-like pulsation occurs there
and the star disappears from optical detection
due to mass loss and the formation of a dense obscuring circumstellar shell.
At this time, the primary mode (orange strip) lies
at the long period edge of sequence C$^{\prime}$.
Related to this is the fact that where the orange channel (primary periods)
crosses sequence C, the secondary periods lie
at the long period edge of sequence C$^{\prime}$.
This suggests an evolutionary sequence during which stars evolve
through the regions of sequences B and C$^{\prime}$
with the 1O mode as the dominant mode of pulsation
and the fundamental mode as a secondary oscillation.
Then at the long period edge of sequence C$^{\prime}$,
the fundamental mode takes over as the dominant mode
and the 1O mode becomes the secondary mode.
But very soon, large amplitude Mira-like pulsation occurs
so that the 1O mode can no longer be detected
and then mass loss occurs and the star disappears from optical detection.


\subsection{An evolutionary picture of pulsational stability}
\label{sec:stability_evolution}

\begin{figure}[ht!]
\epsscale{1.15}
\plotone{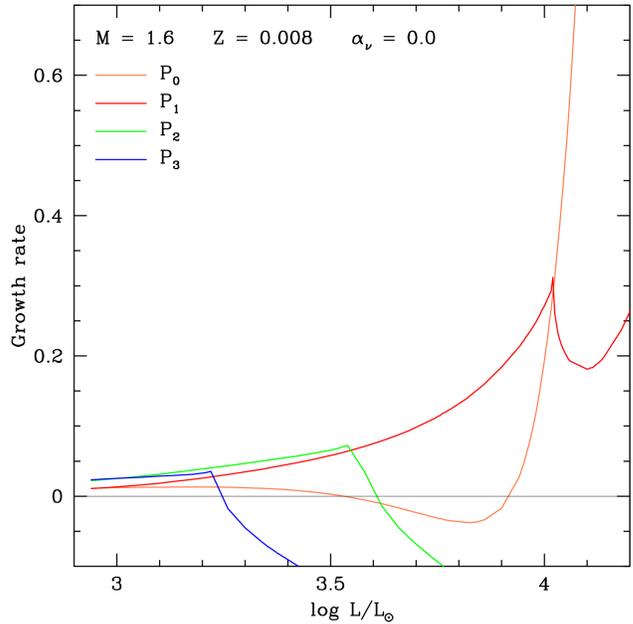}
\caption{
    The growth rate of the first four radial pulsation modes
    in a 1.6 $\Msun$ star as a function of luminosity.
    The metallicity is LMC-like (Z=0.008)
    and turbulent viscosity has been neglected.
    See \citet{WoodOlivier2014} for details of the modelling.
\label{fig:gr-lgl_4modes_alphanu0.0}}
\end{figure}

Theoretical estimates of the stability of radial pulsation modes
in red giant stars support the picture of modal evolution described above.
In Fig.~\ref{fig:gr-lgl_4modes_alphanu0.0},
we show the growth rate of the fundamental, 1O, 2O and 3O radial modes
as a function of luminosity in a 1.6 $\Msun$ star
using the code discussed in \citet{WoodOlivier2014}.

At the lowest luminosities shown, all modes are unstable
with lower order modes tending to have the highest growth rates
and these are presumably the ones that are most likely
to be the detected as the primary period of oscillation.
This accounts for the prominence of sequences A$^{\prime}$, A and B
at low luminosities in Fig.~\ref{fig:wjk-logp_ogleIII_fixed_channels}.
As the luminosity increases, the 3O mode becomes stable
soon after its frequency exceeds the acoustic cutoff frequency
($\log$($L/\Lsun$)$\,\approx 3.22$).
This means that the primary mode of pulsation
will no longer be on sequence A$^{\prime}$
but will lie on sequence A which corresponds
to the 2O mode which now has the highest growth rate. 
Similarly, the 2O mode becomes stable
as the luminosity increases further
($\log$($L/\Lsun$)$\,\approx 3.55$)
so that the primary mode of pulsation now moves to sequence B
which corresponds to the 1O mode which now has the highest growth rate.

The evolution of the 1O mode is somewhat different
to the evolution of the 2O and 3O modes.
With further increase in luminosity,
the 1O mode is never stabilised by
acoustic energy loss in these calculations.
Note that the fundamental mode has become stable
by the time the 1O mode becomes the mode with the highest growth rate.
In lower mass models, the fundamental mode may not be completely stable
but its growth rate is still much smaller than that of the 1O mode.
At very high luminosities, the fundamental mode becomes extremely unstable.  
These results suggest the following picture for further evolution of the star.
The star evolves so that the observed primary mode,
corresponding to unstable 1O mode, 
moves through sequences B and C$^{\prime}$
while at the same time the observed secondary mode,
corresponding to the stable or weakly unstable fundamental mode,
moves across the gap between sequences C$^{\prime}$ and C. 
When the star evolves to the luminosity where the fundamental mode becomes unstable, 
the star becomes a large amplitude Mira-like variable 
and the 1O mode is no longer seen
(defining the long period edge of sequence C$^{\prime}$).
This is followed by mass loss and termination of giant branch evolution,
thus defining the long period edge of sequence~C.

\begin{figure*}[ht!]
\epsscale{1.04}
\plotone{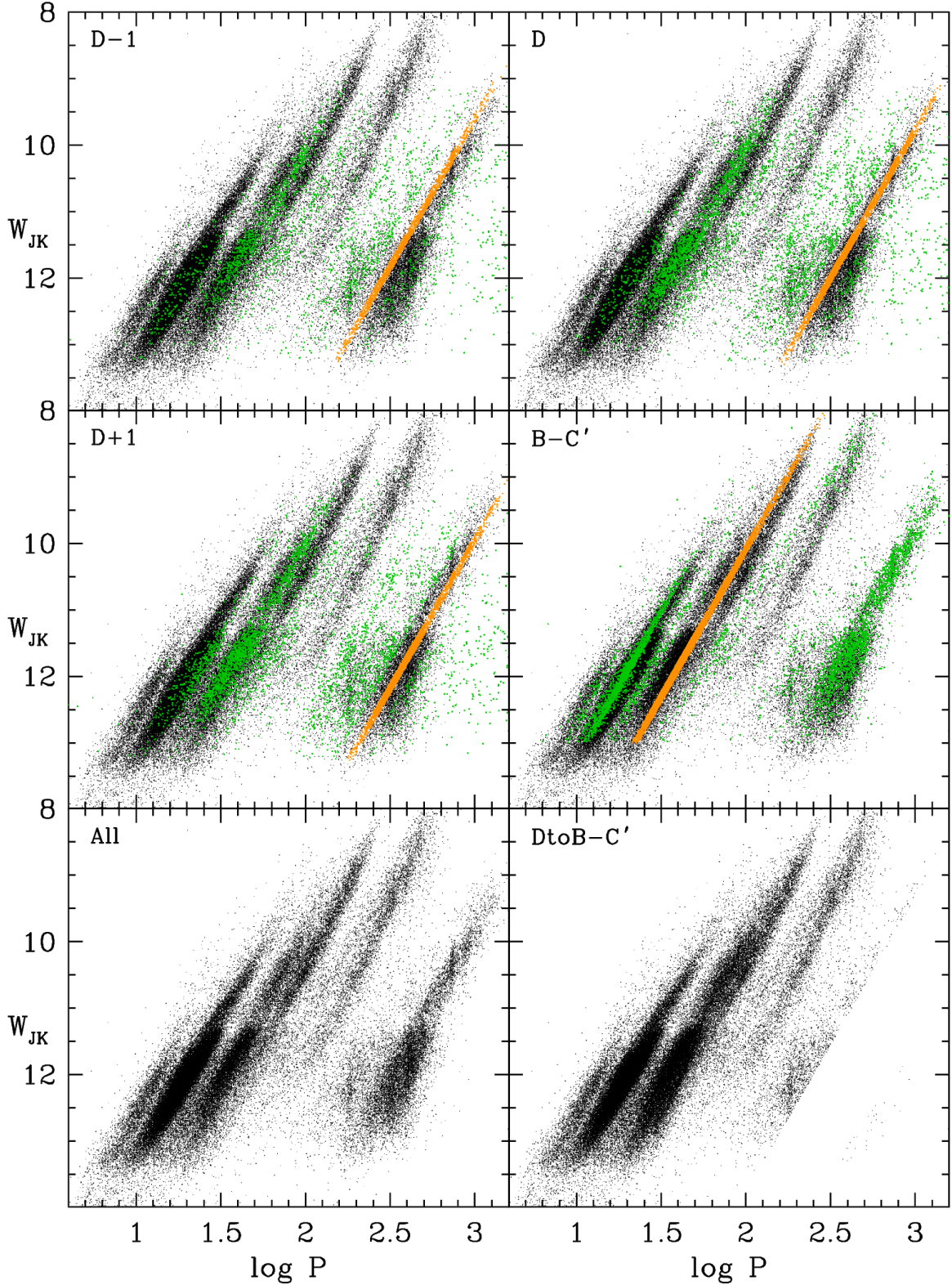}
\caption{
    Similar to Fig.~\ref{fig:wjk-logp_ogleIII_fixed_channels}. 
    In panels (D-1), (D) and (D+1),
    the channels are placed in sequence D.
    In panel (B-C$^{\prime}$),
    the meaning of orange and green points is reversed
    so that the orange points
    correspond to secondary or tertiary periods 
    and the green points are the primary periods
    of the stars marked with orange points.
    In panel (All), the raw OGLE III data is plotted.
    In panel (DtoB-C$^{\prime}$),
    the sequence D variables have been shifted
    to a position lying between sequences
    B and C$^{\prime}$, as described in the text. 
\label{fig:wjk-logp_ogleIII_fixed_channels_seqD}}
\end{figure*}

\begin{figure}[ht!]
\epsscale{1.15}
\plotone{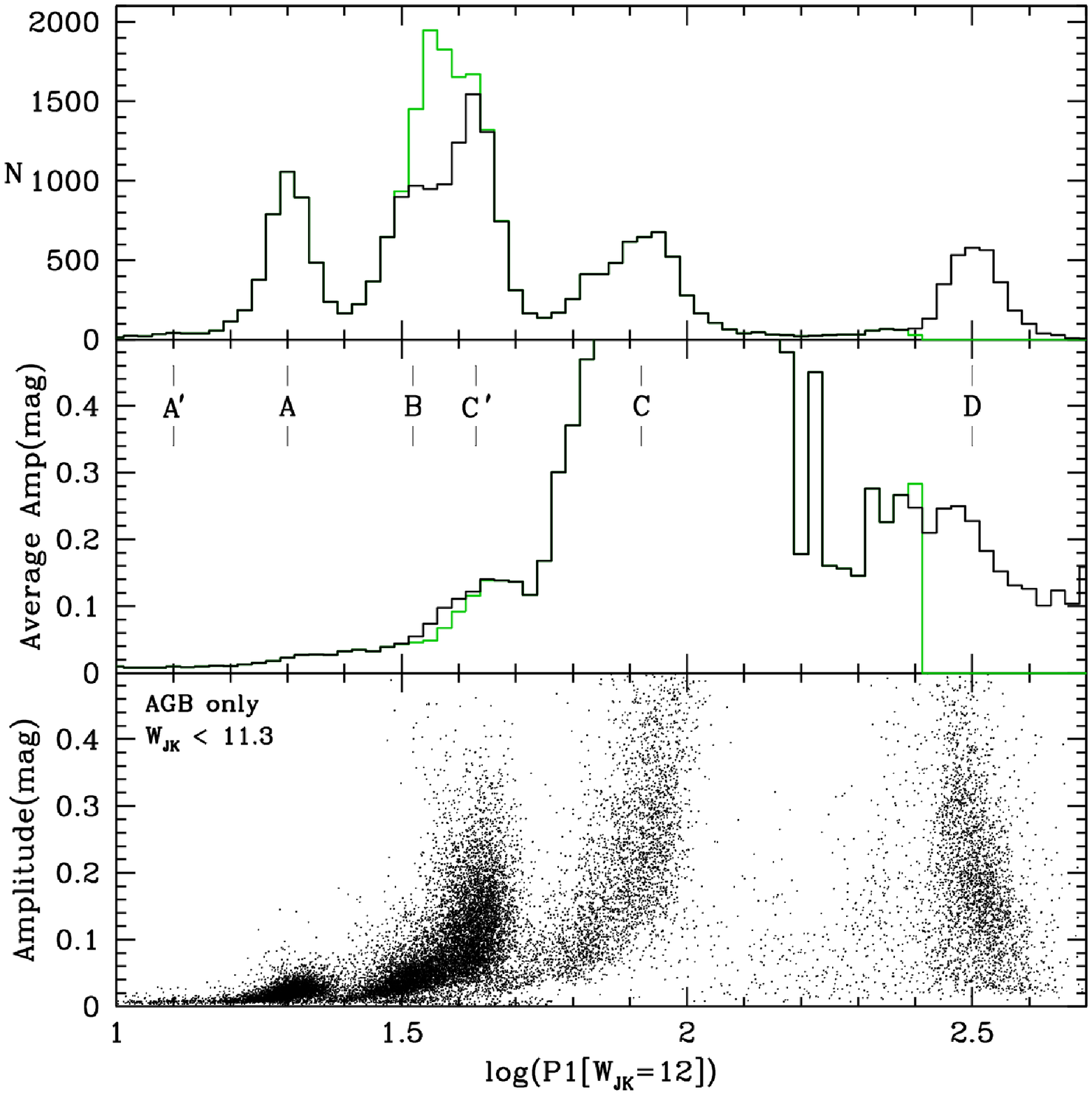}
\caption{
    Bottom panel:
    Pulsation amplitude of individual AGB stars in the OGLE III catalogue
    plotted against $\logPWJKone$.
    The selected AGB stars have $\wjk < 11.3$.
    Middle panel:
    The average amplitude of the AGB stars
    in bins of width 0.025 in $\logPWJKone$.
    Top panel:
    The numbers of stars in the bins. 
    The positions of the PL sequences
    are shown in the middle panel.
    Black lines and points are plotted
    when the stars have the values of $\logPWJKone$
    in the OGLE III catalogue.
    The green lines in the top panel
    are plotted when $\logPWJKone$ values
    for all stars in sequence D have been shifted
    to a new value between sequences B and C$^{\prime}$,
    as described in the text.
    The green lines in the middle panel
    are plotted when only the $40\%$ of sequence D stars
    with a secondary or tertiary period between sequences 
    B and C$^{\prime}$ have been shifted.
\label{fig:lgp1-average_amp_AGB_seqDmoved}}
\end{figure}


\subsection{The splitting of sequences B and C$^{\prime}$}
\label{sec:gap}
The problem with the picture presented so far
is a lack of explanation for the distinct gap
between sequences B and C$^{\prime}$
even though both sequences contain 1O pulsators.
The whole region from the short period edge of sequence B
to the long period edge of sequence C$^{\prime}$
should be populated roughly uniformly in the above picture.
The solution to this problem appears to be provided
by the mysterious sequence D variables.
In the panels of Fig.~\ref{fig:wjk-logp_ogleIII_fixed_channels_seqD},
we show $\log$($P$)-$\wjk$ plots
similar to those in Fig.~\ref{fig:wjk-logp_ogleIII_fixed_channels}.
In panels (D-1), (D) and (D+1), the orange channel
containing the primary mode pulsation
lies on the left, middle and right side of sequence D, respectively.
Keeping to AGB luminosities
where sequences B and C$^{\prime}$ are both well defined,
we see that the secondary modes of the sequence D variables
lie between sequences B and C$^{\prime}$
regardless of the position of the orange channel on sequence D.
In panel (B-C$^{\prime}$) of Fig.~\ref{fig:wjk-logp_ogleIII_fixed_channels_seqD},
we have selected secondary and tertiary periods in a channel
between sequences B and C$^{\prime}$,
and the corresponding primary periods are shown as green points.
It is evident that many of these primary modes
coincide with sequence D variability.
Therefore, there is clearly a one-to-one correspondence
between sequence D variability
and stars with oscillation periods lying between sequences B and C$^{\prime}$.
The tight correlation of the sequence D period
with 1O pulsation period shown by the green and orange strips
in Fig.~\ref{fig:wjk-logp_ogleIII_fixed_channels_seqD}
strongly suggests that sequence D is associated
with some sort of stellar pulsation (e.g. \citealt{Woodetal2004, Saio2015})
and is not caused by a phenomenon such as binarity
(e.g. \citep{Woodetal2004}; \citep{Soszynski_binary2007}).

How the sequence D oscillations are related to 1O pulsation is not clear.
However, it does seem that sequence D variability
is the sole cause of the gap between sequences B and C$^{\prime}$.
Stars that would, in the absence of sequence D variability,
be positioned in the gap between sequences B and C$^{\prime}$,
are now positioned in sequence D.
In the following, we will show that by transferring
all the sequence D stars
into the gap between sequences B and C$^{\prime}$,
the gap disappears.

To begin this process,
we identify all stars having a primary period on sequence D
and a secondary or tertiary period
lying in a strip of width 0.12 in $\log(P)$
located between sequences B and C$^{\prime}$.
These stars, $40\%$ of all stars on sequence D,
are then shifted from the position of their primary period on sequence D
to the position of their secondary or tertiary period
between sequences B and C$^{\prime}$.
The remaining stars in the sequence D region
(60\% of the original sequence D population)
are also shifted into the strip between sequences B and C$^{\prime}$
with an offset in $\log(P)$ from the centre of the strip
equal to half the offset of the sequence D period
from the centre of sequence D.
Now all the original sequence D stars
lie in between sequences B and C$^{\prime}$.
The PL sequences after this shift has been completed
are shown in panel (DtoB-C$^{\prime}$)
of Fig.~\ref{fig:wjk-logp_ogleIII_fixed_channels_seqD}.
These can be compared to the original PL sequences
shown in panel (All) of Fig.~\ref{fig:wjk-logp_ogleIII_fixed_channels_seqD}.

It is clear that after the shifting procedure,
the gap between sequences B and C$^{\prime}$
essentially disappears in the PL diagram.  
This can be seen more clearly by looking at the cumulative numbers
of stars in narrow strips parallel to the sequences.
In Fig.~\ref{fig:lgp1-average_amp_AGB_seqDmoved},
the numbers of AGB stars (defined as those with $\wjk < 11.3$)
in narrow strips of width 0.025 in $\log(P)$
are plotted against $\logPWJK$.
Before the sequence D stars are moved,
there is a dip in the number of stars
in the strips lying between sequences B and C$^{\prime}$
(see the black lines in the top panel
of Fig.~\ref{fig:lgp1-average_amp_AGB_seqDmoved}).
Furthermore, the pulsation amplitude of stars in this region
increases smoothly with $P$
(middle panel of Fig.~\ref{fig:lgp1-average_amp_AGB_seqDmoved})
suggesting that a single pulsation mode is involved.
After moving the sequence D stars,
the whole region on and between sequences B and C$^{\prime}$
is populated as if it is a single sequence
(green lines in the top panel),
once again suggesting that a single mode is involved.
Interestingly, the average amplitude of stars
between sequences B and C$^{\prime}$
is reduced by adding in the $40\%$ of sequence D stars
that have a secondary or tertiary period there.

We conclude that the gap between sequences B and C$^{\prime}$
is actually a selection effect:
the stars in this gap develop a sequence D oscillation
that is of larger amplitude than the sequence B-C$^{\prime}$ oscillation
and hence, in plots using the observed primary period of variability,
these stars appear on sequence D rather than between sequences B and C$^{\prime}$.  

We do not know why the sequence D oscillation occurs.  
It could be due to a resonance between 1O mode pulsation
and the oscillation associated with sequence D
which drains energy from the 1O mode
to the LSP oscillation of sequence D,
but it could equally be a totally independent mode
that just happens to have a large amplitude at this stage of evolution. 


\section{Summary and Conclusions}
\label{sec:conclusions}

We have modelled a synthetic stellar population 
representative of red giant stars currently in the LMC,
and we have assigned to each star a dominant period
based on the growth rate of radial pulsation modes.
We find good agreement between the theoretical PL sequences
formed by the third and second overtone modes when they are dominant
and the observed sequences A$^{\prime}$ and A, respectively.
However, the theoretical first overtone sequence
spreads across both observed sequences B and C$^{\prime}$,
suggesting that these two sequences
may in fact coincide with the same radial pulsation mode.
The theoretical fundamental mode sequence obtained with this modelling
agrees with sequence C at RGB luminosities
but at AGB luminosities it
has periods that are longer than those of sequence C
which contains the large amplitude Mira variables
and which almost certainly does correspond
to radial fundamental mode pulsation \citep{Wood2015}.

The analysis of observed amplitudes
reveals the presence of four distinct period-luminosity-amplitude relations,
that we identify with the four lowest radial orders of pulsation.
The amplitude distributions compare well
with the distributions of theoretical growth rates of overtone modes,
validating the use of the latter as proxies for amplitudes.
However, as with the the PL sequences,
the fundamental mode amplitude distribution
shows a poor agreement with observations,
especially at bright magnitudes.

If we examine where the theoretical fundamental periods fall
regardless of growth rates, we find that they cover the region
from the lower part of sequence B to the entire sequence C.
This is exactly the region covered by observed stars
having modes associated with the observed PAL relation
that we identify as fundamental mode pulsation.
This suggests that while the the growth rates
of the fundamental mode in the models
may be unreliable and underestimated at high luminosities, 
the periods are generally correct.

Using an empirical examination of observations
of multimode LPVs in the LMC,
we have found the following.
Sequence C consists of radial fundamental mode pulsators
which are near the tip of the giant branch
where the fundamental mode has become unstable.
Sequences B and C$^{\prime}$ consist of first overtone mode radial pulsators,
with some fundamental mode pulsators on the lower luminosity end of both sequences.
These fundamental mode pulsators are more massive
than the first overtone mode pulsators
on the same sequence at the same luminosity.

The gap between sequences B and C$^{\prime}$
is caused by the development of a large amplitude oscillation
associated with sequence D
although the nature of this oscillation is unknown.
Sequence A consists of second overtone mode radial pulsators
and sequence A$^{\prime}$ consists of third overtone mode radial pulsators.
Some nonradial modes with angular degree $\ell=1$ and $\ell=2$
have also been detected on sequences A and B, and possibly A$^{\prime}$.

Our two approaches produce quite consistent results
regarding the assignment of modes to the observed PL sequences.
However, since the pulsation-population synthesis models
do not include the mysterious sequence D variability,
they are not able to split the first overtone pulsation sequence
into the two observed sequences.  

Allowing both sequence B and sequence C$^{\prime}$
to correspond to first overtone pulsation
means that the studies of \citet{Mosser2013} and \citet{Wood2015}
can be brought into alignment:
\citet{Mosser2013} can now assign the fundamental mode to sequence C
while \citet{Wood2015} can now move the mode assigned
to sequences A$^{\prime}$, A and B
down by one radial order
to give the same mode assignment as \citet{Mosser2013}.
The major remaining problem with the study of LPVs
is the lack of an understanding of sequence D variability.

\acknowledgments

We acknowledge the support from the ERC Consolidator Grant funding scheme ({\em project STARKEY}, G.A. n. 615604).
This publication makes use of data products from the Two Micron All Sky Survey,
which is a joint project of the University of Massachusetts
and the Infrared Processing and Analysis Center/California Institute of Technology,
funded by the National Aeronautics and Space Administration and the National Science Foundation.

\bibliography{20170830TrabucchiV1}

\end{document}